\documentclass[aps,pra,nofootinbib,preprint,amsmath,amssymb,floatfix]{revtex4}
\usepackage{amssymb}
\usepackage{amsmath}
\usepackage{color}
\usepackage{graphicx}

\begin{document}

\newcommand{\tc}{\textcolor}
\newcommand{\g}{blue}
\newcommand{\ve}{\varepsilon}
\title{  Quantum Electrodynamics and the Ether}         
\author{  Iver Brevik$^1$  }      
\affiliation{$^1$Department of Energy and Process Engineering,  Norwegian University of Science and Technology, N-7491 Trondheim, Norway}

\date{\today}          

\begin{abstract}
The ether concept - abandoned for a long time but reinstated   by Dirac in 1951-1953 - has in recent years emerged into a fashionable subject in theoretical physics, now usually with the name  of the Einstein-Dirac ether. It means that one special inertial frame is singled out, as the "rest frame". What is emphasized in the present  note, is that the idea is a  natural example of the covariant theory of quantum electrodynamics in media if the refractive index  is set equal to unity.    A treatise on this case of quantum electrodynamics was given by the present author back in 1971,  published then only within a preprint series.   The present version is a brief summary  of that formalism, with a link to the original paper. We think it is one of the first treatises on modern ether theory.

\end{abstract}
\maketitle

\bigskip

Consider first the general case where a uniform and isotropic medium with constant refractive index $n=\sqrt{\varepsilon\mu}$ is  moving with constant four-velocity $V_\mu$. We employ coordinates $x_\mu = (x_0,{\bf x})=
(t,{\bf x})$, with $g_{00}=+1$ and put  $\hbar=c=1$, so that $V^\mu V_\mu=1$. This means that one special inertial frame is singled out, namely the medium's rest frame where $V^0=1$.

Assume now that $n=1$, but maintain the overall medium picture. In classical electrodynamics, a free field corresponds to the Lagrangian density
\begin{equation}
{\cal L} = -\frac{1}{4}F_{\mu\nu}F^{\mu\nu}, \label{1}
\end{equation}
where $F_{\mu\nu}=\partial_\mu A_\nu-\partial_\nu A_\mu$. The variational equations obtained from Eq.~(\ref{1}) are
\begin{equation}
\Box{A}_\mu -\partial_\mu \partial^\nu A_\nu=0, \label{2}
\end{equation}
($\Box =\partial^\mu \partial_\mu)$. These are equivalent to Maxwell's equations
\begin{equation}
\partial_\lambda F_{\mu\nu}+\partial_\mu F_{\nu\lambda}+\partial_\nu F_{\lambda\mu}=0, \quad
\partial^\nu F_{\mu\nu}=0. \label{3}
\end{equation}
In quantum electrodynamics, one may  instead start from the Fermi gauge Lagrangian density
\begin{equation}
{\cal L} = -\frac{1}{4}F_{\mu\nu}F^{\mu\nu} - \frac{1}{2}(\partial^\mu A_\mu)^2, \label{4}
\end{equation}
and apply the conventional canonical quantization procedure. The variational equations following from Eq.~({4}) are
\begin{equation}
\Box A_\mu=0, \label{5}
\end{equation}
so that in order to satisfy Maxwell's equations, the Lorentz condition
\begin{equation}
\partial^\mu A_\mu=0 \label{6}
\end{equation}
has to be imposed as an operator condition.

As is known, the Lorentz condition runs into conflict with the canonical commutation rules (for a general discussion, see Ref.~\cite{kallen58}). Maxwell's equations  (\ref{2}) do not fit into the canonical scheme straightaway.  Also, an analysis based upon axiomatic field theory \cite{strocchi70}, confirms this statement.

There exist various ways for how to solve the gauge problem in the quantum case. One approach - the one that we will follow here - is to make use of the special "rest" frame anticipated above, corresponding to $V^0=1$. The Fourier components of the four-potential are expanded into a covariant $k$ dependent basis $e_\mu^{(\lambda)}(k)$, whereby the potential components along the basis vectors become Lorentz invariant quantities. We quantize two of these components and retain the other two as c-numbers. The method has some resemblance with the Coulomb gauge method, although as we work in the Fermi gauge we have the freedom to perform gauge transformations $A_\mu \rightarrow A_\mu+\partial_\mu \chi$, with the condition $\Box \chi =0$. A gauge transformation affects the physically important c-number components an leaves the physically important components unaffected. The Fock space gets a Lorentz invariant meaning, and there is no need of introducing an indefinite metric. The vacuum, defined in a gauge dependent way, is the state where both the number of physical photons and the c-number components are zero. There is some similarity between the present method and the  the Gupta-Bleuler theory \cite{gupta50,bleuler50}; in the latter case the unphysical longitudinal and scalar photons compensate each other.

Turn now to the quantization. For convenience we summarize the following properties of the theory:
\begin{itemize}
\item The potential $A_\mu$ transforms as a four-vector.
\item Equations (\ref{5}) and (\ref{6}), and accordingly also Eq.~(\ref{2}), are satisfied as operator equations.
\item The potential components are decomposed into a covariant basis,
\begin{equation}
a_\mu(k)=
 \sum_{\lambda=0}^3 e_\mu^{(\lambda)}(k)a^{(\lambda)}(k), \label{7}
\end{equation}
where the
$e_\mu^{(\lambda)}$ are covariant vectors containing $k^\mu$ and $V^\nu$. The expressions are complicated and we do not give them here, although we mention that they fulfil the following basic properties,
\begin{equation}
g^{\mu\nu}e_\mu^{(\lambda)} e_\nu^{(\lambda')}= g^{\lambda \lambda'},
 \quad \sum_{\lambda,\lambda'=0}^3 g_{\lambda \lambda'}e_\mu^{(\lambda)} e_\nu^{(\lambda')}= g_{\mu\nu}, \label{8}
\end{equation}
implying that in the rest system, \, ${\bf e}^{(2)} \times {\bf e}^{(3)}= {\bf k}/|{\bf k}|$.
\end{itemize}

We turn now to the commutation relations. Only those components which in the rest system are orthogonal to $\bf k$ are quantized,
\begin{equation*}
[a^{(\lambda)}(k), a^{(\lambda')^\dagger}(k')]=
 \delta_{\bf{k k'}}\delta_{\lambda\lambda'}, \quad \lambda =2,3,
 \end{equation*}
 \begin{equation}
   [a^{(\lambda)}(k), a^{(\lambda')^\dagger}(k')]= \delta_{\bf{k k'}}\delta_{\lambda\lambda'}, \quad \lambda =0,1. \label{9}
  \end{equation}
  The components $a^{(0)}, a^{(1)}$ are c-numbers, and because of the covariance of the vectors $e_\mu^{(\lambda)}$ they also become Lorentz invariants. In view of the Lorentz condition, they satisfy the relation
  \begin{equation}
  a^{(0)} = a^{(1)}. \label{10}
  \end{equation}
  Consider now the projection operator $\tau_{\mu\nu}$ that picks out the transverse part $A_\mu^\perp$ of the potential,
  \begin{equation}
  A_\mu^\perp = \tau_{\mu\nu} A^\nu. \label{11}
  \end{equation}
  It is given by
  \begin{equation}
  \tau_{\mu\nu}= g_{\mu\nu}+\frac{\partial_\mu \partial_\nu}{(V \cdot \partial)^2} -\frac{V_\mu \partial_\nu+V_\nu \partial_\mu}{V\cdot \partial}. \label{12}
  \end{equation}
  Introducing the usual commutator function $D(x)$ and the anticommutator function $D^{(1)}(x)$  \cite{kallen58}, we can write the photon propagator as
  \begin{equation}
  D_{\mu\nu}(x-y)=-\frac{1}{2}\tau_{\mu\nu}D^{(1)} (x-y)-\frac{1}{2}\varepsilon (x-y)\tau_{\mu\nu}D(x-y), \label{12}
  \end{equation}
  where $\varepsilon(x)= x_0/|x_0|$. In Fourier space,
  \begin{equation}
  D_{\mu\nu}(k)= \frac{-i}{k^2+i\eta}\left[ g_{\mu\nu}-V_\mu V_\nu + \frac{(k_\mu-V_\mu \, V\cdot k)(k_\nu-V_\nu \, V\cdot k)}{(V\cdot k)^2-k^2} \right], \label{13}
  \end{equation}
  with $\eta =0^+$.

  \bigskip
  {\bf Discussion}

  The underlying idea of the above formalism is the same as that of relativistic covariant phenomenological electrodynamics in media. There is one  particular inertial frame, called the rest frame, where the formalism simplifies significantly. In the case of a medium with $n>1$, the covariant formalism is constructed such that it reduces to the standard electrodynamic formalism in the rest frame.

  In the formalism above, we set $n=1$, though still maintaining the picture of a preferred inertial frame. As we have seen, the introduction of a preferred frame makes the quantization of the electromagnetic field convenient. The indefinite metric becomes unnecessary.  It lies at hand to associate $V_\mu$ with the four-velocity of an ether, in a modern quantum mechanical sense. In that way $V_\mu$ may be taken to reflect a physical property, instead of being only a formal remedy to simplify the formalism. As mentioned, the ether idea was reinstated again, in a quantum mechanical sense, by Dirac in 1951-1953 \cite{dirac51}.

  The original paper of the author, published in the series "Theoretical Physics Seminar in Trondheim", No. 4, 1971, cannot be uploaded here for technical reasons, but can be assessed via the link
https://hdl.handle.net/11250/2756581 .

\end{document}